# Fermi Surfaces of Surface States on Si(111) + Ag, Au

J. N. Crain[a], K. N. Altmann, Ch. Bromberger[a], and F. J. Himpsel

*Department of Physics, UW-Madison, 1150 University Ave., Madison, WI 53706, USA*
[a] *Department of Physics, Philipps - University, Marburg, Germany*

**Abstract**
Metallic surface states on semiconducting substrates provide an opportunity to study low-dimensional electrons decoupled from the bulk. Angle resolved photoemission is used to determine the Fermi surface, group velocity, and effective mass for surface states on Si(111)√3×√3-Ag, Si(111)√3×√3-Au, and Si(111)√21×√21-(Ag+Au). For Si(111)√3×√3-Ag the Fermi surface consists of small electron pockets populated by electrons from a few % excess Ag. For Si(111)√21×√21-(Ag+Au) the added Au forms a new, metallic band. The √21×√21 superlattice leads to an intricate surface umklapp pattern and to minigaps of 110 meV, giving an interaction potential of 55 meV for the √21×√21 superlattice.

## 1. Introduction

Metallic surface states on semiconductors are special because they are the electronic equivalent of a two-dimensional electron gas in free space. Electronic states at the Fermi level $E_F$ are purely two-dimensional since there are no bulk states inside the gap of the semiconductor that they can couple to. The states near $E_F$ are particularly interesting because they determine most electronic properties, such as conductivity, superconductivity, magnetism, and charge density waves. Recent reports include the effects of charge density waves [1-3], a two-dimensional plasmon in a metallic surface-state band [4], increased surface conductivity via adatom doping [5], and metallic umklapp bands arising from a discommensurate overlayer [6]. In particular, surface conductivity is receiving increased interest with the arrival of sophisticated micro-probes, such as 4-point STM probes [7-11].

Therefore, we have set out to explore the Fermi surfaces that characterize such two-dimensional states using angle-resolved photoemission with an energy resolution better than the thermal broadening of the Fermi edge (3.5 $k_B T$). Angular multi-detection allows us to cover **k**-space with a very fine grid and to resolve electron pockets containing as little as a few percent of an electron per surface atom. That makes it possible to explore the origin and the character of the bands that make the surface metallic. Two mechanisms are found, the first originating from doping by a few percent of extra surface atoms, the other due to a new band associated with a regular array of surface atoms. A complex array of Fermi surfaces is observed which the superlattice forms via surface umklapp.

Metallic surfaces are uncommon on semiconductors, but several structures of noble metals on Si(111) have been discovered recently that are clearly metallic. Here we focus on two-dimensional structures, i.e., Si(111)√3×√3-Ag, Si(111)√3×√3-Au, Si(111)√21×√21-(Ag+Au). Other surfaces with one-dimensional features have been explored separately [12-14].

## 2. Experiment

For preparing surface structures with optimum uniformity and ordering we found the stoichiometry and the annealing sequence to be the most critical parameters. First we will describe calibration of the optimum Au and Ag coverages by low energy electron diffraction (LEED) and the optimum annealing sequences and their rationale. The technical details of the photoemission experiment will be described at the end of this section.

As definition for the Ag and Au coverages we use units of Si(111) atomic layers (1 ML = $7.8 \cdot 10^{14}$ atoms/cm$^2$). The Si(111)√3×√3-Ag structure exists near 1 monolayer Ag coverage. Deposition of slightly over 1 monolayer at 500$^0$C followed by subsequent post-annealing at 500$^0$ C ensures complete coverage with no residual Si(111) 7x7 or Si(111) 3x1- Ag, that occur below the optimum coverages. The Si(111)√3×√3-Au surface comes in several varieties, α-√3×√3 at about 0.9 monolayer coverage, β-√3×√3 at greater than 1.0 monolayer, and 6×6 at 1.1 monolayer. The coverage phase diagrams for the β-√3×√3 and 6×6 overlap and the preparation of the β-√3×√3 differs only in a fast quench from 700$^0$C as opposed to a gradual cooling for 6x6. All three reconstructions exhibit the same basic √3×√3 diffraction spots only with different surrounding fine structure spots. The similarity of the core-level photoemission between the three also suggests a similarity in local structure [15]. In our study we focus on the β and α phases. Both are formed at a large range of annealing temperatures. In order to study a range of intermediate coverages we first deposited a high coverage (1.2 monolayers) and then gradually annealed off Au at 930$^0$C which is close to the Au desorption temperature. The mixed Si(111)√21×√21-(Ag+Au) surface was obtained by evaporating 0.24 monolayer Au at room temperature on top of the 1 monolayer of Ag for the Si(111)√3×√3-Ag surface. Our LEED observations (not shown) are in line with previous work [16, 17].

Fermi surfaces and band dispersions were acquired with a hemispherical Scienta photoelectron spectrometer equipped with angle and energy multi-detection. It was coupled to a 4m normal-incidence monochromator at the Synchrotron Radiation Center (SRC) in Madison. The photon energy hν=34eV was selected, which gave the optimum cross section for surface states in previous work on clean Si(111) and Si(111)-Au [18]. The energy resolution was 29 meV for the photons and 27 meV for the electrons. The light was p-polarized with the plane of incidence along the $[1\bar{1}0]$ azimuth (horizontal in Figs. 1-5). The polar angle between photons and electrons was $50^0$. The polar emission angle of the photoelectrons was varied in steps of $0.5^0$ by rotating the sample with the analyzer staying fixed. This rotation corresponds to the long, horizontal axis of our Fermi surface plots in Figs. 1-5, which is the $[1\bar{1}0]$ component of $\mathbf{k}^\parallel$. The perpendicular $[11\bar{2}]$ direction was covered by multi-detection over a $12^0$ range with the data points $0.24^0$ apart. Scans were taken with an energy window of about 1eV in 0.005eV steps.

In this fashion we acquire the photoemission intensity in a three-dimensional parameter space, i.e., the energy E and the two in-plane k-components $k_{[1\bar{1}0]}$ and $k_{[11\bar{2}]}$. By taking slices at a constant energy $E = E_f$ we plot the Fermi surfaces (Figs. 1-5). Cuts at constant $k_{[1\bar{1}0]}$ map E versus k dispersions along $[11\bar{2}]$ (Figs. 6-7). The original coordinates are the polar angle $\vartheta_{[1\bar{1}0]}$ and the multi-detection angle $\vartheta_{[11\bar{2}]}$. These angles are converted to their corresponding parallel wave vectors according to

$$\mathbf{k_x} = \mathbf{k}^\parallel_{[1\bar{1}0]} = \hbar^{-1}(2mE_{kin})^{1/2} \cdot \cos\vartheta_{[11\bar{2}]} \cdot \sin\vartheta_{[1\bar{1}0]}$$
and
$$\mathbf{k_y} = \mathbf{k}^\parallel_{[11\bar{2}]} = \hbar^{-1}(2mE_{kin})^{1/2} \cdot \sin\vartheta_{[11\bar{2}]}.$$

Here $\vartheta_{[11\bar{2}]}$ is always smaller than $\pm 6^0$, and for simplicity we use the approximation $\cos\vartheta_{[11\bar{2}]} \approx 1$ in Figs. 1-5 with an error < 1%.

The individual scans were normalized according to a procedure outlined in a previous report [18]. Before measurement, the alignment and angular transmission function of the multidetector were carefully adjusted to provide homogeneous intensity across the $12^0$ angular range. Any residual inhomogeneities are removed by taking an additional set of reference scans at the same spectrometer settings as the data, but with an 8eV higher photon energy. At this energy we expect to see only secondary electrons with little angular dependence. The reference scans are integrated over energy and then used to divide out the transmission function in the raw data. Each scan, including references, is normalized to the beam current to account for the steady decline in the incident photon flux. The areal photon density within the acceptance spot of the spectrometer changes slowly with the angle of incidence of the light, causing a slow variation in the intensity that is roughly proportional to $\cos\vartheta$. The intensity of each scan was adjusted to account for this factor.

Our geometry allows us to cover most of the 1×1 surface Brillouin zone and nearly three √3×√3 zones (Fig. 1b), with a grid dense enough to resolve fine structure within the small √21×√21 Brillouin zone (Fig. 3). Even though the E(**k**) band dispersions along the $k_x$ and negative $k_x$ directions are symmetric, the photon polarization is not. Along $k_x$ the component perpendicular to the surface dominates, whereas along negative $k_x$ the in-plane component dominates. That allows us conclusions about the symmetry of the surface states using polarization selection rules.

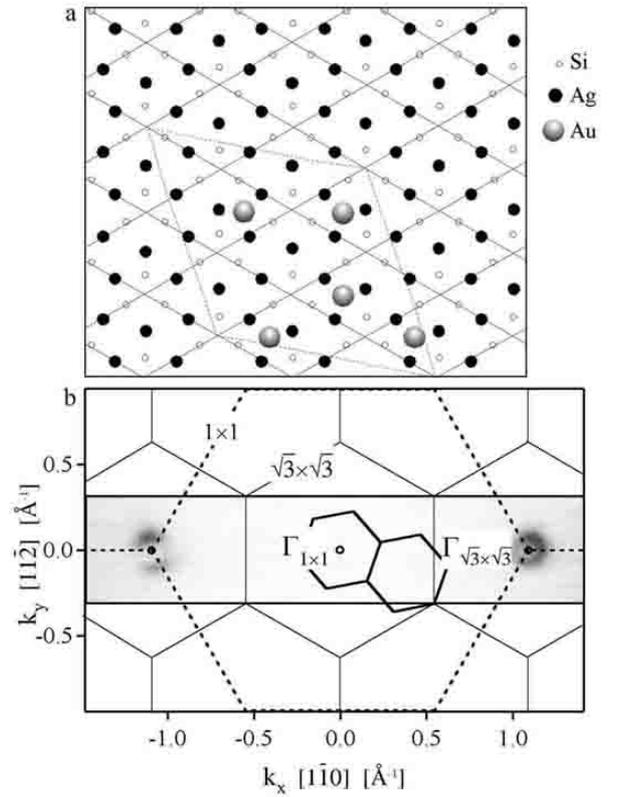

**Fig. 1** Two-dimensional surface structures of Ag and Au on Ag(111) in real and reciprocal space. a) Structural models of Si(111)√3×√3-Ag and Si(111)√21×√21-(Ag+Au). The latter is obtained by adding 5/21=0.24 monolayer Au to the √3×√3-Ag surface. b) Surface Brillouin zones for the Si(111) 1×1, √3×√3, and √21×√21 structures. The experimental Fermi surface of Si(111)√3×√3-Ag is superimposed, which consists of two small circles around the $\Gamma_{\sqrt{3}\times\sqrt{3}}$ points at $k_x = \pm 1.1 \text{Å}^{-1}$.

In Fig. 1 we plot a variety of unit cells together, both in real space (top) and reciprocal space (bottom). The real space √3×√3 schematic is of the HCT model for Ag [19]. The extra Au atoms show a possible model for the Si(111)√21×√21-(Ag + Au) surface that is consistent with recent x-ray diffraction data [17, 20]. The structure for the √3×√3 Au - α phase is similar to that of the √3×√3 Ag (For the differences see [21-23]). The wave vectors of the various symmetry points are: $\overline{K}_{1\times1} = 2\sqrt{2}/3 \cdot 2\pi/a = 1.09$ Å$^{-1}$; $\overline{M}_{1\times1} = \sqrt{2}/\sqrt{3} \cdot 2\pi/a = 0.95$ Å$^{-1}$; $\overline{K}_{\sqrt{3}\times\sqrt{3}} = 0.63$ Å$^{-1}$; $\overline{M}_{\sqrt{3}\times\sqrt{3}} = 0.55$ Å$^{-1}$. The reciprocal lattice points of the √21×√21 structure correspond to the center of the circles in Fig. 3c.

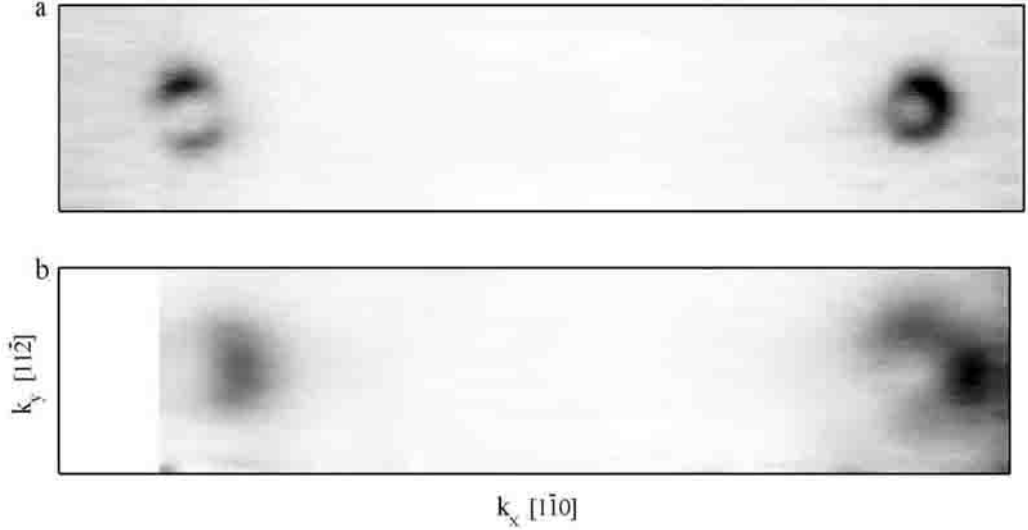

**Fig. 2** Two-dimensional Fermi surfaces (= lines) obtained from the photoemission intensity at $E_F$. High intensity is shown dark, **k**=0 is at the center of the upper frame (compare Fig. 1). The electric field vector **E** is close to the x-direction on the left and close to the z-direction on the right. a) Si(111)√3×√3-Ag exhibits two small circles around the $\Gamma_{\sqrt{3}\times\sqrt{3}}$ points. The circle on the left is split by a node in intensity due to polarization selection rules. b) Si(111) α √3×√3 – Au exhibits two small electron pockets similar to those with Ag but with different intensities, indicating different symmetry.

### 3. Fermi Surfaces and Band Dispersion

This section presents the Fermi surfaces in Figs. 2-5 and the E(**k**) band dispersions in Figs. 6,7 for each of the three surfaces studied. The bands are mapped along the $k_y$ direction ( = [ 1 1 $\overline{2}$ ] ) using angular multidetection, which allows us to resolve small Fermi surfaces and band gaps.

### 3.1 Si(111)√3×√3-Ag

Figures 1b and 2a map the Fermi surface for Si(111)√3×√3-Ag, using an energy integration interval of ±0.025 eV around $E_F$. High photoemission intensity is shown dark. The √3×√3 Fermi surface consists of a single circle of radius k = 0.076 Å$^{-1}$ that is centered at the extra √3×√3 Γ points (Fig. 1b). The 1×1 Γ point at k = 0 exhibits no detectable trace of an equivalent circle, in agreement with a previously identified surface state labeled $S_1$[24-27]. Most of the unmapped portions of the 1$^{st}$ 1x1 Brillouin zone can be inferred by symmetry.

Previous photoemission reports on the Si(111)√3×√3-Ag surface have painted a complicated picture of polarization dependent effects [24, 26, 28]. By mapping the entire Fermi surface in 2D instead of just a single line, we derive a more complete picture of these matrix-element effects. In Figs. 1b, 2a one observes an asymmetry between positive and negative $k_x$ that is due to polarization dependent effects. For negative $k_x$ the angle of incidence of the light is nearly normal ($\theta_i = 27^0$, k = -1.09Å$^{-1}$) and the resulting polarization vector is in the plane of the sample along [ 1 $\overline{1}$ 0 ]. In this case the emission intensity has a node in the x-direction. This matrix element effect agrees with the original findings [26] while recent work [28] finds that the metallic surface state $S_1$ exhibits no measurable polarization dependence. This can perhaps be understood because of the narrowness of the missing intensity around the $k_x$ symmetry line in Figure 2. A deviation of only a few degrees from the ( $\overline{1}$ $\overline{1}$ 2 ) plane might disguise the polarization dependence. Ref. [28] also reports that the $S_1$ state is excited only by the component of the photoelectric vector parallel to the surface, i.e., the state consists mainly of $p_x$ and $p_y$ components [24, 28]. However, we clearly observe the state for grazing incidence of the light ($\theta_i = 73^0$ corresponds to $k_x = 1.09$Å$^{-1}$ in Fig. 2a) with the polarization nearly perpendicular to the surface. This apparent discrepancy is likely due to the different photon energies used, 21.2eV and 34eV respectively.

Past studies have found that the perpendicular components of surface states on Si(111) and on Si(111) with noble metal adsorbates are excited preferentially at 34eV and are much weaker at 21eV [12, 13, 18, 29]. Thus it is not surprising that in previous studies at 21.2eV the surface state was hardly visible for grazing incidence of the light.

## 3.2 Si(111)√21×√21-(Ag+Au)

The evaporation of only 0.24 monolayers of additional Au on the √3×√3-Ag surface yields a Si(111)√21×√21-(Ag+Au) superstructure that exhibits new pieces of the Fermi surface. These are shown in Fig. 3a (at $E_F$ – 0.1 eV) and in Fig. 4b (at $E_F$) with an energy integration interval of ±0.025 eV. The two prominent features are a large outer circle that is concentric with a small inner circle (Fig. 4 b,c). The two Fermi surfaces that we observe are consistent with band dispersions from earlier photoemission studies of Si(111) √21×√21-(Ag+Au) which report outer and inner bands, labeled $S_1^*$ and $S_1'$ respectively [24, 30] (for the $E_k$ comparison see Figure 7). In Fig. 4a and 4b we compare the Fermi surfaces near the Γ point from Figs. 2b and 3. The similarity of the inner pocket of 4b with 4a suggests that the inner band $S_1'$ of the √21×√21-(Ag+Au) is actually a vestige of $S_1$ of the √3×√3-Ag, while the new band $S_1^*$ arises from the additional Au adatoms. Indeed, after a gentle anneal (Figs. 4c) the inner band $S_1'$ appears identical to the √3×√3-Ag. While these strong features are centered at the √3×√3 Γ point, fainter circles fill the rest of the Brillouin zone. To better observe these faint bands, the Fermi surface is re-plotted using a logarithmic scale in Fig. 3b. The resulting pattern of circles is obtained by translating the strongest outer circle by reciprocal lattice vectors of the √21×√21 lattice i.e., via surface umklapp (Fig. 3c).

In addition to the unit cell shown in Fig. 1a there is a second (mirror) domain rotated at an angle of 21.8⁰ with respect to the first [25], which has been included in Fig. 3a. The result is in excellent agreement with the experimental Fermi surface. The umklapp features have much lower intensity than the original circle, and there is a rapid decrease in intensity away from the √3×√3 Γ point. Previous studies have observed a similar effect in LEED images of the Si(111) √21×√21-(Ag+Au) surface where the √21×√21 spots that surround the √3×√3 spots have much stronger intensity than those surrounding the integer order spots [17]. We might expect similar effects to be observed in photoemission since it is a time-reversed LEED process.

The association of $S_1^*$ with the extra Au atoms is supported by the umklapp bands (Fig. 3) which duplicate $S_1^*$ with the periodicity of the √21×√21 reciprocal lattice vectors. In contrast, √21×√21 umklapps of $S_1'$ are absent. The only umklapp of $S_1'$ that does appear is at the 1×1 Γ point which corresponds to a shift by a √3×√3 reciprocal lattice vector. It is interesting that the small pocket is observed at $Γ_{1×1}$ with extra Au while it is absent with Ag only.

Since $S_1^*$ is directly linked to the extra Au atoms, simple electron counting arguments can be used to test the coverage of Au and the number of bands in the √21×√21 unit cell. The radius of the Fermi surface is k = 0.266 ± 0.008 Å$^{-1}$ which means that $S_1^*$ subtends an area of π × (0.266 Å$^{-1}$)$^2$ = 0.22 ± 0.01 Å$^{-2}$. This is 0.072 times the area of the 1×1 Brillouin zone (3.09 Å$^{-2}$) and 1.51 the area of the √21×√21 Brillouin zone.

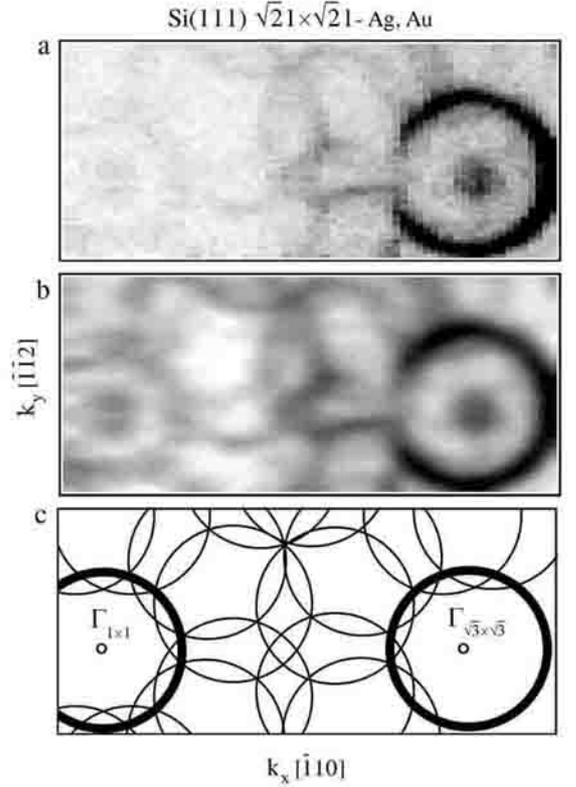

**Fig. 3** Energy surface at $E_F$–0.1eV for Si(111)√21×√21-(Ag+Au). The small electron pocket of the √3×√3-Ag substrate (dark dot) is surrounded by a large, dark circle originating from the electrons of the additional Au atoms. Fainter circles originate from surface umklapp. a) Raw data with high photoemission intensity shown dark. b) Smoothened data on a logarithmic gray scale, showing fainter umklapp features. c) Schematic of the umklapp features, obtained by shifting the bold circles by reciprocal lattice vectors of the √21×√21 structure.

This area corresponds to 3.0 ± 0.2 electrons per √21×√21 unit cell using Luttinger's theorem and assuming a single band containing electrons with both spins. Since each Au atom contributes one unpaired s,p electron we can compare this number with the actual Au coverage, which is 0.24 monolayer in our experiment, or 5 atoms per √21×√21 unit cell. That is consistent with a recent X-ray diffraction study [20] and a STM study [17] which support the model with 5 Au atoms per √21×√21 unit cell shown in Fig. 1a. The observed extra band can only accommodate 3.0 of the 5 electrons in a naïve single band interpretation. Clearly, one has to go beyond that. Two possibilities are discussed in Section 4. A pair of degenerate bands could accommodate 6.0 electrons, and an extra, filled band at lower energy would increase the electron count to 5.0. The second option clearly appears closer to the experimental coverage, although we cannot completely rule out the degenerate band option.

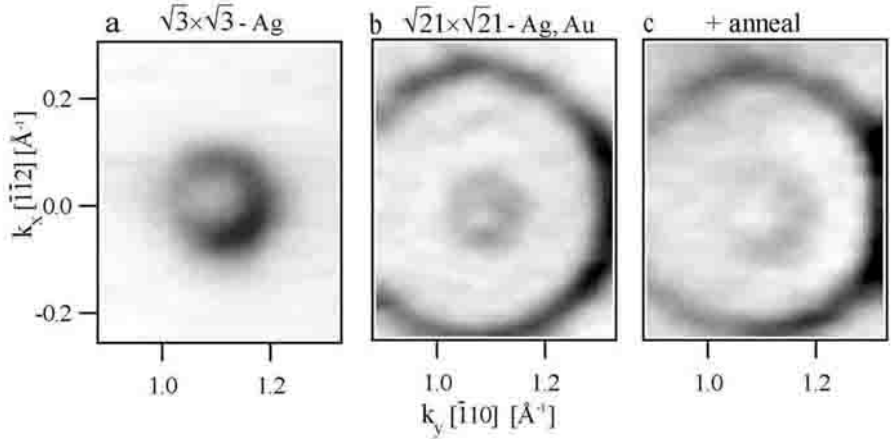

**Fig. 4** Close-up of the Fermi surface around the $\Gamma_{\sqrt{3}\times\sqrt{3}}$ point (compare Fig. 1). a) Si(111)$\sqrt{3}\times\sqrt{3}$-Ag with a coverage slightly larger than a monolayer, giving rise to small electron pockets for the excess electrons. b) After depositing 0.24 monolayer Au onto a) to produce Si(111)$\sqrt{21}\times\sqrt{21}$-(Ag+Au). c) After annealing b) to 300$^0$C for 20 seconds. The central Ag-induced pocket widens to its original size.

The small electron pocket at the center of the large $\sqrt{21}\times\sqrt{21}$-(Ag+Au) pocket has not always been observed in previous studies [24, 31]. There are three possible reasons for such an absence. 1) The central pocket is only seen with a few % of excess Ag on top of the $\sqrt{3}\times\sqrt{3}$-Ag structure [32] which forms the starting point of the $\sqrt{21}\times\sqrt{21}$-(Ag+Au) surface. In fact our observed polarization dependence of the small pocket is identical to that on $\sqrt{3}\times\sqrt{3}$-Ag, indicating that it has the same origin. 2) Polarization selection rules quench emission when the **E** vector is parallel to $k_x$ (= [1 $\bar{1}$ 0]), as in [26, 31]. 3) The Au coverage is too low (0.15 ML in [31]).

**Table 1** Band minimum $E_0$, Fermi wave vector $k_F$, effective mass $m_{eff}$, and group velocity $v_F$ of metallic surface states on Si(111).

| Surface | $E_0$ (eV) | $k_F$ (Å$^{-1}$) | $m_{eff}$ | $v_F$ ($10^6$m/s) |
|---|---|---|---|---|
| Si(111)$\sqrt{3}\times\sqrt{3}$-Ag[*] 1+ε monolayer[**] Ag | −0.32 | 0.08 | 0.07 | 1.3 |
| Si(111)$\sqrt{21}\times\sqrt{21}$- (Ag+Au) 1 monolayer Ag + 0.24 monolayer Au | −0.78[***] | 0.27 | 0.35 | 0.9 |
| Si(111)$\sqrt{3}\times\sqrt{3}$-Au α: 0.9 monolayer Au | −0.32 | 0.15 | 0.25 | 0.7 |

---

[*] Previous work [24] reported $E_0$= −0.18eV, $k_F$=0.11Å$^{-1}$, and $m_{eff}$ = 0.25. The bottom of the band $E_0$ lies higher than in our work because of less additional Ag, corresponding to lower band filling. The larger effective mass is probably due to the use of energy distribution curves (EDCs), where the Fermi cutoff makes it difficult to determine the band position near $E_F$. In our study we use momentum distribution curves (MDCs) to determine the bands, which avoids this problem.

[**] ε ≈ 0.01 estimated from the area inside the Fermi surface, assuming a single, non-degenerate band [19, 22, 36].

[***] Ref. [24] reported $E_0$= −0.62eV and $m_{eff}$ = 0.29. Here again the Au and Ag coverages are lower, explaining the higher $E_0$ by lower band filling.

The band dispersion for the Si(111)$\sqrt{21}\times\sqrt{21}$-(Ag+Au) surface is shown in Fig. 7a,b together with a model calculation in c). The umklapp bands show up as faint lines in a) and b) and are represented as dashed lines in c). An additional dashed parabola represents the central pocket left over from the $\sqrt{3}\times\sqrt{3}$-Ag starting surface. The Brillouin zone boundary of the $\sqrt{21}\times\sqrt{21}$ lattice is at $k_y$=0.210 Å$^{-1}$, which coincides with the intersection of the strong primary bands with the umklapp bands. The band dispersion deviates from a straight line, showing a vertical displacement at the zone boundary. This distortion can be explained by a mini-gap induced by the periodic potential of the $\sqrt{21}\times\sqrt{21}$ lattice. The gap can be quantified using the energy distribution curves in Fig. 7d, which are vertical cuts in Figs. 7 a for three momenta close to the zone boundary ZB. The bands avoid crossing each other and form a double-peak with equal intensities right at the zone boundary, instead of a single peak (Fig. 7d middle). After determining the peak positions from a fit with constrained, equal widths for all curves we find a gap energy $E_g$ = 110 meV. It is directly related to an interaction potential $U = E_g/2$ = 55 meV.

### 3.3 Si(111)$\sqrt{3}\times\sqrt{3}$-Au

In Figure 2b the Fermi surface for Si(111) α $\sqrt{3}\times\sqrt{3}$ - Au is plotted. The region in k-space is identical as for that of the $\sqrt{3}\times\sqrt{3}$ - Ag allowing for direct comparison. As with the Ag, the Fermi surface of α $\sqrt{3}\times\sqrt{3}$ - Au has a single pocket centered at the $\sqrt{3}\times\sqrt{3}$ Γ point. This corresponds to a metallic surface state reported in earlier photoemission and inverse photoemission studies [29, 33, 34]. The exact shape of the Fermi surface is difficult to determine because of its asymmetry across the Γ point in both the x and y directions, and the general broadness of the band. These asymmetries must be related to the dipole matrix element.

Of the Si(111) $\sqrt{3}\times\sqrt{3}$ - Au family of phases, the α is most similar to the $\sqrt{3}\times\sqrt{3}$ – Ag. Both occur near 1 ML of coverage (0.9 ML for Au) and have some similarity in LEED and STM. However, X-ray diffraction studies of the $\sqrt{3}\times\sqrt{3}$ – α Au reveal a conjugate honeycomb-chained-trimer structure (CHCT) that represents a reversal of the silicon and noble metal atoms as compared to the honeycomb-chained-trimer structure (HCT) for √3×√3 – Ag [21-23]. At higher Au coverages the more complicated β √3×√3 and 6×6 phases have no direct analog in the √3×√3 Ag system. Each of the Au phases shares a similar in local atomic structure, and it has been proposed that they are constructed of tilings of such structural units. The 6×6 is built of an ordered tiling of trimer and pentagonal Au units [35]. The β phase is similar to the 6×6 phase except for the lack of long range order. In the lower coverage α phase the trimer tiling predominates as the CHCT model suggests. It has been proposed that the pentagonal unit, or incomplete versions of it, becomes more prominent at higher coverages. In this manner the reconstruction can accommodate a wide range of Au coverages, which explains the observed wide range of coverages, from 0.8 to 1.5 ML.

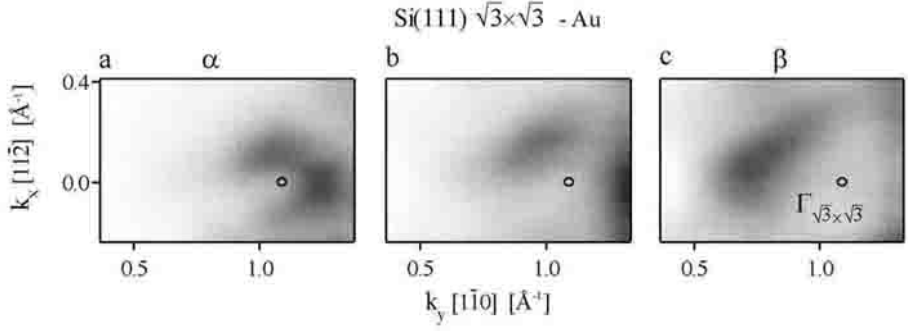

**Fig. 5** The Fermi surface for Si(111)√3×√3-Au with the coverage increasing from a) to c). A continuous opening of the pocket at $\Gamma_{\sqrt{3}\times\sqrt{3}}$ is observed, in contrast to the extra Fermi surface that is created when adding Au to the √3×√3-Ag surface.

Observations of the Fermi surface as a function of the Au coverage are consistent with this concept. In Figure 5 we map the Fermi surface near the √3×√3 Γ point for three different coverages of Au. Fig. 5a corresponds to 0.9 ML Au (α phase), Fig. 5c corresponds to 1.2 ML (β phase), and Fig. 5b corresponds to an intermediate coverage. The size of the pocket at the Γ point increases as a function of the Au coverage. This contrasts to the addition of Au to the √3×√3 – Ag surface, for which a new band was observed. The increase in area is attributed to the addition of electrons to the unit cell accompanying the extra Au. The band also becomes very broad, especially for the intermediate Au coverage, which is consistent with the lack of long-range order for the higher Au coverages.

### 4. Discussion

The Fermi surface data of Si(111)√3×√3-Ag in Figs. 1b, 2a provide not only the band filling from the area of the Fermi surface (Table 1), but also shed light on the wave function symmetry via the pronounced polarization dependence.

The filling of the surface state band was attributed to excess Ag in a recent core level and ARPES study on Si(111)√3×√3-Ag. It was found that the surface state band is empty at the optimum Ag coverage of 1 monolayer [32]. The surface state band became filled when additional Ag atoms were present in small quantities beyond the stoichiometric coverage. Our preparation, in which more than 1ML of Ag is deposited and the excess is annealed off, is consistent with some filling of the surface state by excess Ag. If we assume a single band (with two spins) crossing the Fermi level, as in first principles calculations [22, 36], the area inside the Fermi surface gives an electron count of 0.01 electrons per 1×1 unit cell, which would correspond to an excess of 0.01 monolayer Ag.

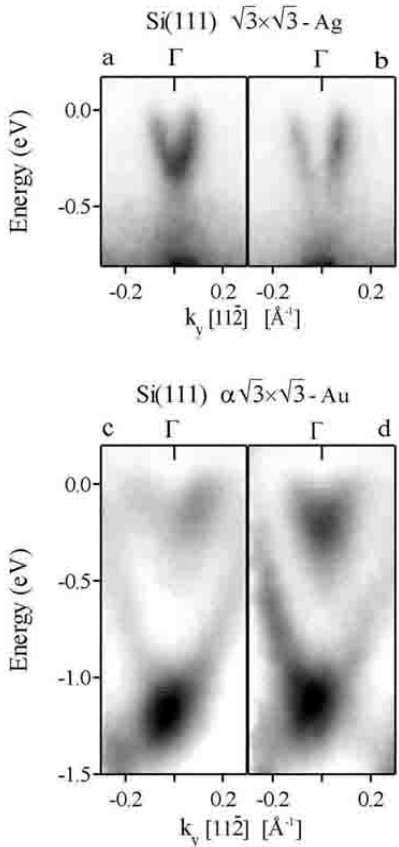

**Fig. 6** E($\mathbf{k}$) band dispersions obtained by multi-detection (high photoemission intensity is dark). All panels are cuts along $k_y$ through the pockets at $\Gamma_{\sqrt{3}\times\sqrt{3}}$ (see Fig. 1). $E_F = 0$. a) Si(111)√3×√3-Ag for $k_x= +1.1\text{Å}^{-1}$. b) Si(111)√3×√3-Ag for $k_x= -1.1\text{Å}^{-1}$, showing the node at $k_y=0$. c) Si(111)α√3×√3-Au for $k_x= +1.1\text{Å}^{-1}$. The outer band does not reach the Fermi level. d) Si(111)α√3×√3-Au for $k_x= -1.1\text{Å}^{-1}$. The node at $k_y=0$ is absent.

The surface state band is absent at the central Γ point, which is common to the 1×1 and √3×√3 lattice (see Fig. 1b). This observation is consistent with earlier work on Si(111) √3×√3 –Ag [26]. Such behavior can be explained by coupling with bulk states at the valence band maximum, which is located at this Γ point. The other two Γ points on the left and right are characteristic of the √3×√3 lattice and fall into a gap of bulk states. Surface umklapp would be required to bring bulk states from $k_{x,y}=0$ to these two Γ points.

A strong polarization dependence has been observed previously, and it is clarified in our data by observing the states in the $k_x,k_y$ plane, not only along the $k_x$ or $k_y$ direction. A node is observed in the photoemission intensity along the $k_x$ direction when the polarization vector **E** is parallel to x = [ 1 $\bar{1}$ 0 ] (left side of Figs. 1b,2a). Such a node in a high symmetry plane is reminiscent of dipole selection rules for a mirror plane, although one has to keep in mind that the y=0 plane corresponds to ( $\bar{1}$ $\bar{1}$ 2 ), which is not a mirror plane. For a mirror plane the node would indicate out-of-plane $p_y$ orbital character because odd states can only be excited with the **E**-vector perpendicular to the plane, as long as the emission is within the plane. The node is absent on the right side of Figs. 1b, 2a, where the **E**-vector is in the plane and nearly perpendicular to the surface. Such behavior would be consistent with $p_z$ orbital character (or possibly $p_x$). A naive combination of the two results seems to indicate a doubly-degenerate band, but that is not supported by first principles calculations [22, 36]. Another alternative might be a single $p_{y+z}$ band. While such assignments are still rather speculative, it is clear that the polarization data provide a critical test of possible wave functions. To fully exploit them requires quantitative studies of the photoemission matrix element for the two geometries.

The results from Si(111)√21×√21-(Ag+Au) exhibit a striking new Fermi surface that can be attributed to the √21×√21 superlattice of adsorbed Au atoms. Replicas of the Au-induced electron pocket induced by translation with √21×√21 reciprocal lattice vectors form a complex maze of Fermi surfaces. These surface umklapp features may be viewed as photoelectrons diffracted by the Au superlattice, analogous to the extra superlattice spots seen in low energy electron diffraction (LEED). The superlattice is also felt in the E(k) band dispersions, where it induces mini-gaps at the zone boundaries. These mini-gaps are directly correlated to the superlattice potential, thereby providing a quantitative information about the interactions of the electrons with the superlattice (55 meV in this case)

The measured Fermi surface area of the Si(111)√21×√21-(Ag+Au) surface can be combined with the electron count to reveal some unexpected electronic features. The observed size of the large electron pocket can only explain 3 of the 5 electrons introduced into the √21×√21 unit cell by the 5 adsorbed Au atoms. Assuming that the existing coverage model is correct, that leaves two possible interpretations. The electron pocket could be doubly-degenerate and thus accommodate 6 electrons rather than 3. Alternatively, a second completely filled band could accommodate the remaining 2 electrons. Our data strongly favor the second model, but we cannot rule out the first model altogether. The extra band accommodating 2 more electrons could easily fit into a region of high density of states observed in the energy region between –0.9 eV and –1.5 eV.

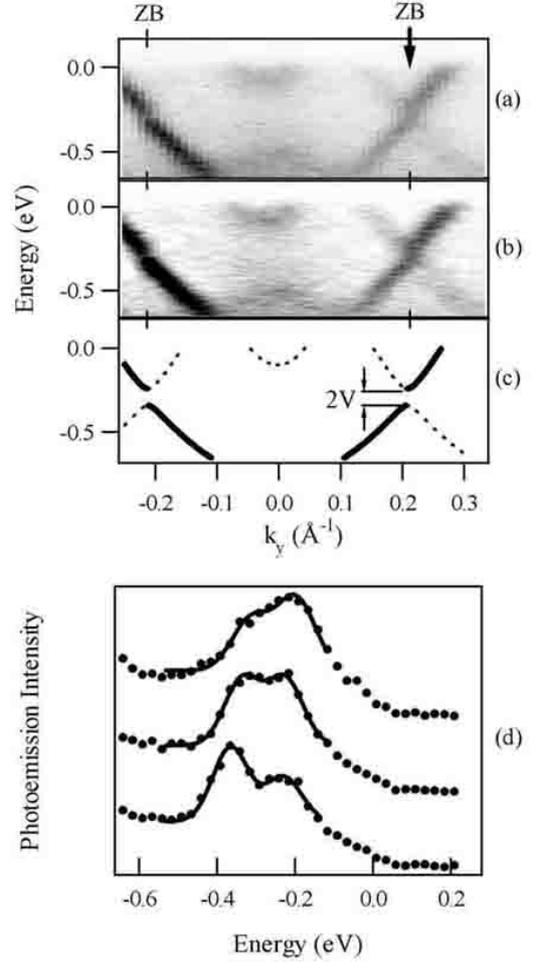

**Fig. 7** E(**k**) band dispersions for Si(111)√21×√21-(Ag+Au). The strong outer band reaches the Fermi level and gives rise to the large Fermi circles in Figs. 3,4. Faint √21 umklapp bands (dashed in c) correspond to the faint circles in Fig. 3. a) Raw data with high photoemission intensity shown dark. b) Smoothened data on a logarithmic gray scale, showing fainter umklapp features and the central √3×√3-Ag pocket. c) Two band model of the Au-induced bands using a √21×√21 interaction potential V = 55meV. Energy distributions corresponding to vertical cuts in a) at the √21×√21 zone boundary ZB (middle curve) and on either side of it (top, bottom). An avoided crossing with a mini-gap of 110 meV is observed.

The Si(111)α√3×√3-Au phase is similar to Si(111)√3×√3-Ag, with both exhibiting electron pockets that are centered around the √3×√3 Γ points that coincide with the 1×1 K points, but not at the Γ point **k**=0. This degree of electronic similarity is somewhat surprising considering their different atomic structures. Likewise, adding Ag to √3×√3-Ag and Au to α√3×√3-Au leads to an increase of the electron pockets which accommodates the extra electrons.

The addition of Au to the √3×√3-Au surface at elevated temperature causes a restructuring of the Au atoms in the surface layer [35]. The local structure remains the same, but some of the predominant trimer units are replaced by pentagonal units to accommodate the additional Au. These structural changes are manifest in an enlargement of the Fermi surface with coverage as well as an increased broadness of the bands.

## 4. Summary

In summary, we have used recent developments in energy and angle multidetection to map the complete, two-dimensional Fermi surfaces of several metallic surface structures induced by Ag and Au on Si(111). The results show that silicon surface states can have well-defined Fermi surfaces despite a tendency to form localized surface orbitals with correlation gaps [37-39]. The metallic states at the Fermi level are well-suited to study low-dimensional electrons because they are decoupled from the bulk in an absolute band gap, and not just in certain $k^{\parallel}$ regions as on metal surfaces.

Using the coverage dependence of the Fermi surfaces we are able to distinguish two general mechanisms for surface metallicity - expansion of an existing Fermi surface by doping and creation of a new Fermi surface. The doping mechanism acts at low excess coverages of Ag and Au (few % of a monolayer). An electron pocket gradually opens up and expands by filling an unoccupied band with electrons from added noble metal atoms. At an excess coverage of 24% of a monolayer, which is characteristic of the √21×√21 structure, a completely new Fermi surface appears.

The extra reciprocal lattice vectors of the √21×√21 superlattice form replicas of the √3×√3 Fermi surface and lead to an intricate web of overlapping Fermi surfaces. The corresponding avoided crossings between surface umklapp bands can be mapped accurately enough to extract the superlattice potential acting on the surface electrons. Such a rich band topology promises interesting new ways of tailoring electronic states at semiconductor surfaces.

**Acknowledgments**: We acknowledge D. Fick for stimulating discussions and support by the University Marburg. We are grateful to M. Bissen, G. Rogers, Ch. Gundelach, and M. Fisher for help with the experimental setup. This work was supported by the NSF under Award Nos. DMR-9704196, DMR-9815416, and DMR-0079983 and by the DOE under Contract No. DE-FG02-01ER45917. It was conducted at the SRC, which is supported by the NSF under Award No. DMR-0084402.